\documentclass{article} 
\usepackage{iclr2025_conference,times}

\usepackage{amsmath,amsfonts,bm}

\def\eqref#1{equation~\ref{#1}}

\def\1{\bm{1}}

\DeclareMathAlphabet{\mathsfit}{\encodingdefault}{\sfdefault}{m}{sl}
\SetMathAlphabet{\mathsfit}{bold}{\encodingdefault}{\sfdefault}{bx}{n}

\usepackage{hyperref}
\usepackage{url}

\title{Rethinking Feature Fusion in Multimodal \\ CTR Prediction }

\author{Junjie Zhou  \\
National Key Laboratory for Novel Software Technology, Nanjing University, China\\
School of Artificial Intelligence, Nanjing University, China\\
\texttt{zhoujj@lamda.nju.edu.cn} \\
}

\iclrfinalcopy 
\begin{document}

\maketitle

\begin{abstract}
With the rapid advancement of Multimodal Large Language Models (MLLMs), an increasing number of researchers are exploring their application in recommendation systems. However, the high latency associated with large models presents a significant challenge for such use cases. The EReL@MIR workshop provided a valuable opportunity to experiment with various approaches aimed at improving the efficiency of multimodal representation learning for information retrieval tasks. As part of the competition's requirements, participants were mandated to submit a technical report detailing their methodologies and findings. Our team was honored to receive the award for Task 2 — Winner (Multimodal CTR Prediction). In this technical report, we present our methods and key findings. Additionally, we propose several directions for future work, particularly focusing on how to effectively integrate recommendation signals into multimodal representations. The codebase for our implementation is publicly available at: \url{https://github.com/Lattice-zjj/MMCTR_Code}, and the trained model weights can be accessed at: \url{https://huggingface.co/FireFlyCourageous/MMCTR_DIN_MicroLens_1M_x1}.
\end{abstract}

\section{Introduction}
Multimodal Large Language Models (MLLMs) represent a transformative paradigm in artificial intelligence, enabling the integration of diverse data modalities—such as text, images, and audio—into unified representations. This advancement has opened new avenues for enhancing recommendation systems, which traditionally rely on unimodal data sources. By leveraging the rich contextual information from multiple modalities, MLLMs have the potential to significantly improve the personalization and accuracy of recommendations.

Recent research in multimodal recommender systems (MRS) has focused on developing models that effectively fuse information from various modalities to capture user preferences more comprehensively. For instance, \cite{liu2024multimodal} categorize existing MRS approaches into four key areas: modality encoding, feature interaction, feature enhancement, and model optimization. These components work in tandem to process and integrate multimodal data, addressing challenges such as modality imbalance and semantic alignment. 

Despite these advancements, deploying MLLMs in real-world recommendation scenarios presents challenges, notably the high computational cost and latency associated with processing large-scale multimodal data. To mitigate these issues, researchers have explored efficient representation learning techniques. For example, \cite{liu2022disentangled} propose a disentangled multimodal representation learning framework that captures users' modality preferences across different factors, enhancing recommendation performance while reducing computational overhead .

In this technical report, we present our approach to multimodal click-through rate (CTR) prediction, developed in the context of the Efficient Representation Learning for Multimodal Information Retrieval (EReL) workshop~\cite{fu20251sterelmirworkshopefficient}. Our method achieved first place in Task 2 (Multimodal CTR Prediction), demonstrating the effectiveness of our strategies in integrating multimodal signals for improved predictive accuracy. We detail our model architecture, feature fusion techniques, training methodologies, and experimental findings. Furthermore, we discuss potential future directions, particularly focusing on incorporating recommendation-specific inductive biases into multimodal representations to better align with user intent and preference modeling.

\section{Related Work}

\subsection{Feature Fusion in Recommender Systems}

Feature fusion is a pivotal aspect of modern recommender systems, particularly with the advent of multimodal data sources such as text, images, and audio. Traditional approaches often employed early fusion techniques, where features from different modalities are concatenated before being input into models. However, this method may not effectively capture the complex interactions between modalities.

Recent studies have explored more sophisticated fusion strategies. For instance, \cite{zhou2023comprehensive} provide a comprehensive survey on multimodal recommender systems, categorizing existing approaches into modality encoding, feature interaction, feature enhancement, and model optimization. They emphasize the importance of effectively integrating multimodal information to improve recommendation performance.

Furthermore, attention-guided multi-step fusion networks have been introduced to model latent item-item semantic structures and optimize item representations through attention-guided strategies~\cite{zhou2023attention}. These approaches highlight the trend towards more dynamic and context-aware fusion techniques in recommender systems.

\subsection{Multimodal Embedding Strategies}

The integration of multimodal embeddings has shown promise in capturing the complementary information present in different data modalities. A study by \cite{liu2022disentangled} introduces a disentangled multimodal representation learning framework that captures users' modality preferences across different factors, enhancing recommendation performance while reducing computational overhead.

In practice, various strategies have been employed to combine textual and visual embeddings. For example, applying Principal Component Analysis (PCA)~\cite{zhou2025all} separately to BERT-based text embeddings and CLIP-based image embeddings before concatenation has been found to preserve modality-specific information more effectively, leading to improved recommendation accuracy.

\section{Methodology \& Experiment}
\subsection{Model Architecture}

Our model builds upon the Deep Interest Network (DIN) framework, integrating several enhancements to effectively capture complex user-item interactions:

\begin{itemize}
    \item \textbf{Embedding Layer}: A unified embedding layer projects both user and item features into a shared latent space, facilitating the modeling of interactions across different feature types.
    \item \textbf{Multi-Head Target Attention}: To capture dynamic user interests, we employ a multi-head attention mechanism that focuses on relevant parts of the user's historical behavior in relation to the target item.
    \item \textbf{Squeeze-and-Excitation Network (SENet)}: Inspired by FiBiNET, we incorporate a SENet module to recalibrate channel-wise feature responses, enhancing the model's ability to emphasize informative features and suppress less useful ones.
    \item \textbf{Deep Neural Network (DNN)}: The concatenated features from the embedding and attention layers are passed through a multi-layer perceptron to capture high-order feature interactions and produce the final prediction.
\end{itemize}

\subsection{Feature Interaction Enhancements}

To further enrich the representation of user-item interactions, we experimented with additional modules:

\begin{itemize}
    \item \textbf{Cross Network}: We integrated a cross network to model explicit feature interactions by applying cross terms between user and item features. This allows the model to capture multiplicative relationships that are not easily learned by standard DNNs.
    \item \textbf{Bilinear Interaction (FiBiNET-inspired)}: We explored bilinear feature interactions as proposed in FiBiNET, enabling the model to learn pairwise feature interactions more effectively by considering the importance of each feature pair.
\end{itemize}

\subsection{Multimodal Item Embeddings}

Recognizing the importance of multimodal information in recommendation systems, we evaluated different strategies for incorporating textual and visual features of items:

\begin{itemize}
    \item \textbf{V1}: PCA-reduced text embeddings derived from BERT representations.
    \item \textbf{V2}: PCA-reduced image embeddings obtained from CLIP representations.
    \item \textbf{V3}: PCA applied to the concatenation of BERT and CLIP embeddings.
    \item \textbf{V4}: Concatenation of separately PCA-reduced BERT and CLIP embeddings.
\end{itemize}

\subsection{Experimental Results}

We conducted extensive experiments to assess the impact of the aforementioned enhancements on model performance. The evaluation metric used was the Area Under the ROC Curve (AUC). The results are summarized in Table~\ref{tab:auc_results}.

\begin{table}[h]
\centering
\caption{AUC Scores for Different Embedding Strategies}
\label{tab:auc_results}
\begin{tabular}{l c}
\hline
\textbf{Embedding Strategy} & \textbf{AUC} \\
\hline
V1: PCA(BERT) & 0.9247 \\
V2: PCA(CLIP) & 0.9231 \\
V3: PCA(BERT + CLIP) & 0.9124 \\
V4: PCA(BERT) + PCA(CLIP) & \textbf{0.9306} \\
\hline
\end{tabular}
\end{table}

The results indicate that the V4 strategy, which involves concatenating separately PCA-reduced BERT and CLIP embeddings, achieves the highest AUC. This suggests that preserving distinct modality characteristics through separate dimensionality reduction before concatenation is more effective than combining modalities prior to PCA.

\section{Future work}
Building upon our current framework, we identify several avenues for future research aimed at enhancing the robustness and effectiveness of multimodal recommendation systems.

\subsection{Contrastive Learning with User-Perceived Similarity}

To better capture user preferences, we propose leveraging downstream recommendation signals to identify items that users perceive as similar. Utilizing techniques such as SWinG, we can extract these similarities and employ contrastive learning methods to train models that more effectively integrate multimodal information. This approach aims to align the learned representations with user perceptions, thereby improving recommendation accuracy.

\subsection{Advanced Quantization Techniques for Feature Extraction}

Our current implementation employs Principal Component Analysis (PCA) for dimensionality reduction of multimodal embeddings. While PCA offers computational efficiency, it may not fully capture the complex structures inherent in multimodal data. Inspired by recent advancements in vector quantization \cite{liu2024vector}, we plan to explore residual quantization and product quantization methods. These techniques have demonstrated superior performance in preserving information while reducing dimensionality, which could lead to more expressive and compact feature representations in our recommendation models.

\subsection{Enhancing Data Quality through Filtering Mechanisms}

During data preprocessing, we observed that some entries, particularly those translated from Chinese to English using Large Language Models (LLMs), contain significant errors, including mistranslations and malformed JSON structures. Incorporating such flawed data into the training set can adversely affect model performance. To address this, we intend to implement data filtering strategies that detect and exclude erroneous entries. Techniques such as influence functions \cite{lam2022analyzing} can be employed to assess the impact of individual data points on model predictions, allowing for the identification and removal of detrimental samples. Additionally, integrating validation checks during data ingestion can further ensure the integrity of the training dataset.

\section{Conclusion}
By integrating contrastive learning based on user-perceived similarities, adopting advanced quantization methods for feature extraction, and enhancing data quality through rigorous filtering, we aim to develop a more robust and effective multimodal recommendation system. These future directions hold the potential to significantly improve the alignment between model predictions and user preferences, thereby enhancing overall recommendation performance.



\bibliography{iclr2025_conference}
\bibliographystyle{iclr2025_conference}


\end{document}